\documentclass[pre,twocolumn,superscriptaddress,showpacs,%
preprintnumbers,amsmath,amssymb]{revtex4}

\usepackage{amsmath}
\usepackage{graphicx}
\usepackage{dcolumn}

\begin{document}

\title{Hysteresis and re-entrant melting of a self-organized \\ system of classical particles confined in a parabolic
trap}

\author{F.~F.~Munarin}
\email{munarin@fisica.ufc.br} \affiliation{Departamento de F\'isica,
Universidade Federal do Cear\'a, Caixa Postal 6030, Campus do Pici, 60455-760 Fortaleza, Cear\'a, Brazil}%
\affiliation{Department of Physics, University of Antwerp, Groenenborgerlaan 171, B-2020 Antwerpen, Belgium}
\author{K.~Nelissen}
\email{kwinten.nelissen@ua.ac.be} \affiliation{Department of Physics, University of Antwerp, Groenenborgerlaan 171,
B-2020 Antwerpen, Belgium}
\author{W.~P.~Ferreira}
\affiliation{Departamento de F\'isica,
Universidade Federal do Cear\'a, Caixa Postal 6030, Campus do Pici, 60455-760 Fortaleza, Cear\'a, Brazil}%
\author{G.~A.~Farias}
\affiliation{Departamento de F\'isica,
Universidade Federal do Cear\'a, Caixa Postal 6030, Campus do Pici, 60455-760 Fortaleza, Cear\'a, Brazil}%

\author{F.~M.~Peeters}
\email{francois.peeters@ua.ac.be} \affiliation{Department of Physics, University of Antwerp, Groenenborgerlaan 171,
B-2020 Antwerpen, Belgium}
\date{ \today }

\begin{abstract}
A self-organized system composed of classical particles confined in a two-dimensional parabolic trap and interacting
through a potential with a short-range attractive part and long-range repulsive part is studied as function of
temperature. The influence of the competition between the short-range attractive part of the inter-particle
potential and its long-range repulsive part on the melting temperature is studied. Different behaviors of the
melting temperature are found depending on the screening length ($\kappa$) and the strength ($B$) of the attractive
part of the inter-particle potential. A re-entrant behavior and a thermal induced phase transition is observed in a
small region of ($\kappa,B$)-space. A structural hysteresis effect is observed as a function of temperature and
physically understood as due to the presence of a potential barrier between different configurations of the system.
\end{abstract}



\maketitle

\section{Introduction}

The study of the properties of self-organized systems has increased dramatically in recent years. This interest
originates from the possibility to control the formation of patterns having an important impact on applications that
use large-scale self-assembly to create specific pattern morphologies. This kind of structures occur in systems from
diverse areas including chemistry \cite{Chem2} and biology \cite{Chang00}. In physics, it was predicted that systems
which fall into this morphological category are generated by the competition between short-range attraction and
long-range repulsion \cite{Sear98}. This competitive interaction appears in many systems, such as magnetic materials
\cite{Stojkovic99,Stojkovic00}, colloids \cite{Bubeck99,Leiderer98}, and two-dimensional electron systems
\cite{Fradkin99, Schmalian00}. Experimentally, colloidal systems are one of the most studied systems, which in
combination with theoretical predictions, may lead to the design of novel soft materials and to an understanding of
the glass and gel state of matter \cite{Sciortino02}. Recently, it was observed that a similar type of pattern
formation can also arise in particular classes of ultrasoft colloids with a strictly repulsive inter-particle
potential \cite{Mladek06}.



Besides presenting a rich variety of cluster types and showing an excellent model for technological applications,
colloidal systems have the added advantage of the facility to control the interaction between particles and of real
time imaging of their configuration through video microscopy. A wide variety of studies of colloidal systems were
performed in order to understand the structure and dynamics of different kind of systems, such as colloidal
particles interacting through a short-range attractive and long-range repulsive potential
\cite{Campbell05,Lu06,Shevchenko06}. For instance, a binary system of superparamagnetic colloidal particles that are
confined by a two-dimensional (2D) water-air interface and exposed to an external magnetic field perpendicular to
the interface showed diverse stable configurations \cite{Hoffmann06}. Moreover, the authors observed that clustering
appeared only for one type of particles, instead of both types of particles.

Recently, several models were developed having a small number of interacting particles in order to understand the
behavior of colloidal systems as a function of temperature \cite{Bedanov94,Kong03,Peeters00}. Different melting
scenarios were studied extensively in systems consisting of charged particles for a range of different
inter-particle interaction and trap types, such as a system of binary charged particles confined by a circular
hard-wall potential interacting by a repulsive dipole potential \cite{Kwinten_Re}, confined by a parabolic trap
potential and interacting by a Coulomb inter-particle potential \cite{Felipe_M}, and non-confined particles with
short-range attraction and long-range repulsion interaction \cite{Reich_InfBuble}.

Motivated by the increased interest in the behavior of systems of particles that are characterized by a competition
between short-range attraction and long-range repulsion, we analyze here the melting of a system composed of a
finite number of classical particles interacting through a potential which is composed of a repulsive Coulomb and an
attractive exponential term. The particles move in a 2D plane that are confined by a parabolic trap. The zero
temperature configurations of this system were studied by Nelissen {\it et al.} \cite{Kwinte_Bubles}. They observed
several kinds of topological different configurations (e.g., ring and bubble configuration). But from this rise the
question about the stability of those kinds of configuration against thermal fluctuation. This motivated us to study
the melting of these ordered configurations and to analyze the effect of the interplay between the short-range and
the long range interaction on the melting process. We found, that if one increases the temperature, some of the
bubble configurations exhibit a thermally induced structural phase transition and a remarkable re-entrant behavior.
In addition, we found that this system exhibits hysteresis behavior for the mean radial displacement when we
increase and decrease the temperature and the configuration goes through a thermally induced structural phase
transition.

This paper is organized as follows. In Sec. II, we describe the mathematical model and the numerical approach to
obtain the ground state configurations. In Sec. III we analyze the melting process for different inter-particle
potentials. Temperature induced re-entrant and hysteresis effects are discussed in Sec. IV. Our conclusions are
presented in Sec. V.

%

\section{Numerical approach}
\label{sec:model}

We consider a 2D cluster with $N$ classical particles interacting through a potential composed of a repulsive
Coulomb and an attractive exponential term as in Ref.~\cite{Reich_InfBuble,Kwinte_Bubles}. The particles are kept
together by a parabolic potential centered at the origin. The general dimensionless Hamiltonian of the system is
written as:
\begin{equation}\label{hamiltonianII}
\begin{split}
  H = \sum_{i=1}^{N}r_i^2 + \sum_{i>j=1}^{N} \left(\frac{1}{\vert \vec{r}_{i} -
  \vec{r}_{j}\vert} - B e^{-\kappa\vert \vec{r}_i - \vec{r}_j\vert}\right),
\end{split}
\end{equation}
where $r_{i}\equiv |\mathbf{r}_i|$ is the distance of the $i$th particles from the center of the parabolic
confinement. The energy and the distance are in units of $E_{0}=(m\omega_{0}^{2}r_{0}^{2}/2)$ and $r_{0}=(2q^{2}/m
\epsilon \omega_{0}^{2})^{1/2}$, respectively. Notice that the $B$ and $\kappa$ parameters determine the exponential
contribution in the hamiltonian, where $B$ determines the strength and $1/\kappa$ the interaction range of the
attractive part of the inter-particle potential (the third term in Eq. (\ref{hamiltonianII})). Note also that the
$\kappa$ parameter is inversely proportional to the range of the attractive part in the potential. Now the state of
the system is determined by $B$, $\kappa$ and the number of particles $N$. Temperature is expressed in units of
$T_{0}=E_{0}/k_{B}$, where $k_{B}$ is the Boltzmann constant.

The ground state configurations ($T=0$) of the two-dimensional system were obtained by the Monte-Carlo (MC)
simulation method (using the standard Metropolis algorithm \cite{Metropolis}) extended with the Newton optimization
method \cite{Peeters95}. The particles were allowed to reach a steady state configuration after $10^{5}$ simulation
steps, starting from different initial random positions. In the same time, we calculate the frequencies of the
normal modes of the system using the Householder diagonalization technique \cite{Peeters95}. The configuration was
taken as final if all frequencies of the normal modes were positive and the energy did not decrease further.

In order to understand the trajectory of each particle correctly, we study the melting of the system using Molecular
Dynamic simulation (MD). The temperature of system was increased from $T=0$ (ground state configuration) with
successive steps of $\delta T$ and equilibrating at the new temperature during $10^6$ MD steps, (with a typical step
size of $\Delta t=0.001$). After this equilibrium, the average energy was calculated, together with the mean squared
radial displacement given by
\begin{equation}\label{radialdeviation}
\displaystyle{
  \langle u_{R}^{2} \rangle \equiv
  \frac{1}{N}\sum_{i=1}^{N}(\langle r_{i}^{2} \rangle - \langle r_{i}
  \rangle ^{2}) / \rho^{2},
}
\end{equation}
where $\rho$ is the average inter-particle distance at zero temperature. The symbol $\langle\rangle$ stands for an
average over typically $10^{6}$ - $10^{7}$ MD steps after equilibration of the system.


The melting temperature was determined through a Lindemann-like criterion, which has been widely used for 2D finite
size clusters. This criterion states that melting occurs when $\langle u_{R}^{2} \rangle$ reaches 0.1 of the
inter-particle distance at zero temperature \cite{lozovik85}. But what is essential is that melting is characterized
by a rapid increase of the fluctuation of particles when temperature reaches the melting temperature.

\section{Melting}


\begin{figure*}
\begin{center}
\includegraphics[scale=0.65]{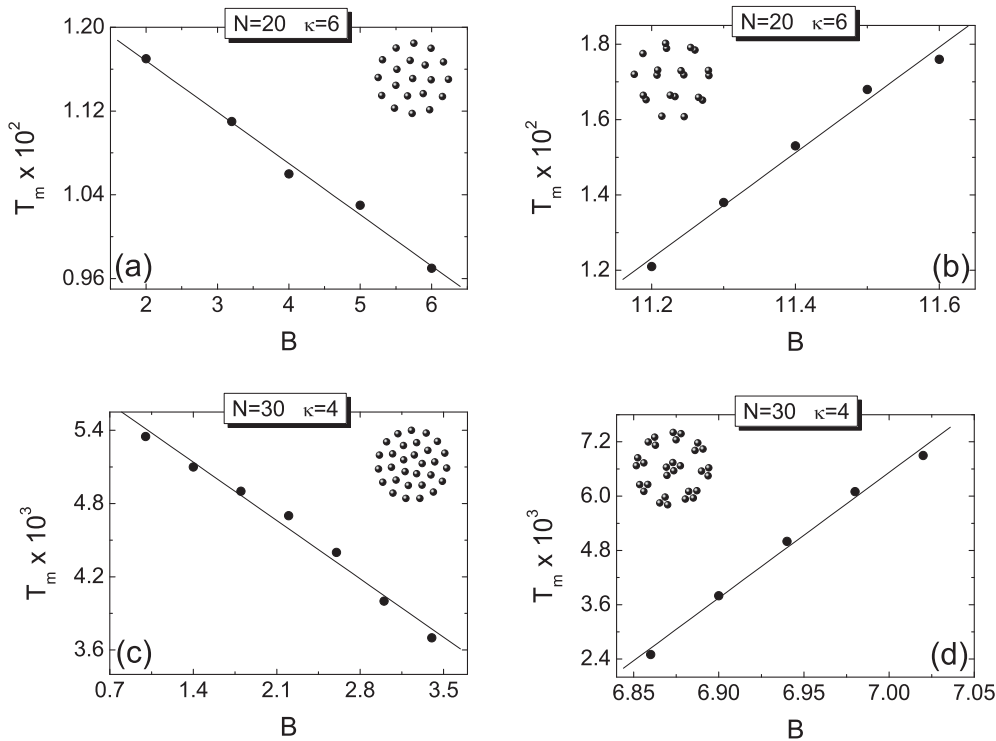}
\caption{The total melting temperature ($T_m$) as a function of $B$ for a system with $\kappa=6$, $N=20$ particles
(a) and (b), and a system with $\kappa=4$, $N=30$ particles (c) and (d). We considered ring configurations (a) and
(c), and bubble configurations (b) and (d). Symbols are the numerical results and the solid line is a guide for the
eye.}\label{fig:Tmelt_VarB}
\end{center}
\end{figure*}

In the following, we analyze the melting temperature and as an example we consider systems composed of $20$ and $30$
particles and study its dependence on the $B$ and $\kappa$ parameters. These two systems are the typical examples of
the $T=0$ configurations consisting of rings or bubbles.

\subsection{Dependence on B}

In this section, we study the melting temperature, as a function of the strength of the attractive part of the
inter-particle interaction ($B$), for $N=20$ and $N=30$ particles and a fixed value of $\kappa$. We observe from
Fig.~\ref{fig:Tmelt_VarB} that different $B$-regions exhibit a different melting temperature ($T_m$) dependence. For
instance, in the case of $N=20$ and $\kappa=6$, we find that the melting temperature decreases when $B$ increases
(see Fig. \ref{fig:Tmelt_VarB}(a)), whereas the opposite behavior is found for large $B$ values (see Fig.
\ref{fig:Tmelt_VarB}(b)). The behavior of Fig. \ref{fig:Tmelt_VarB}(a) is in some sense a surprising result since
$B$ is the strength of the short-range attraction, and one may expect that the larger the attraction strength, the
more packed the particles are, and therefore the higher the melting temperature. This concept is correct since the
attractive part of the potential is large enough to compete with the repulsive potential part, but on the contrary,
the opposite effect happens because the total inter-particle potential is purely repulsive in the case of the
cluster of Fig. \ref{fig:Tmelt_VarB}(a). In this situation, when the value of $B$ increases, the repulsive potential
decreases [see Eq.~(\ref{hamiltonianII})] in the same way as the melting temperature due to the decrease of the
coupling among the particles. This behavior was also observed for the system with $30$ particles and $\kappa=4$
[Fig. \ref{fig:Tmelt_VarB}(c)].

In Figs. \ref{fig:Tmelt_VarB}(b) and \ref{fig:Tmelt_VarB}(d), we show the dependence of the melting temperature as a
function of $B$ for the cluster with $N=20$, $\kappa=6$ and $N=30$, $\kappa=4$, respectively in the large B-region.
As seen, the attractive part of the potential is large enough to form small clusters, and therefore the melting
temperature increases with $B$ as a consequence of the increase of the attractive part of the inter-particle
potential. Hence when the system is composed of bubbles, the melting temperature increases with $B$ due to the
increase of the coupling among particles in the small bubbles. The behavior of Figs. \ref{fig:Tmelt_VarB}(a),
\ref{fig:Tmelt_VarB}(c) and Figs. \ref{fig:Tmelt_VarB}(b), \ref{fig:Tmelt_VarB}(d) clearly shows the dependence of
the melting temperature on $B$, which is opposite for pure ring configurations and bubble configurations.

\begin{figure*}
\begin{center}
\includegraphics[scale=0.65]{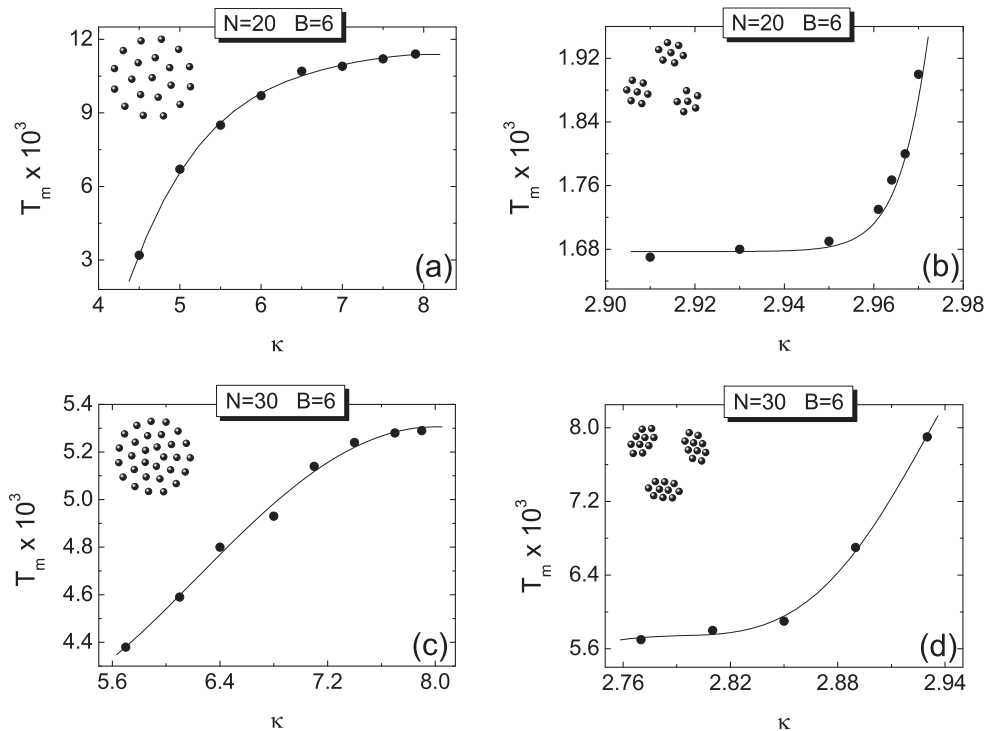}
\caption{ The total melting temperature ($T_m$) as a function of $\kappa$ for a system with $B=6$, $N=20$ particles
(a) and (b), and $N=30$ particles (c) and (d). We considered ring configurations (a) and (c), and bubble
configurations (b) and (d). Symbols are the numerical results and the solid line is a guide for the
eye.}\label{fig:Tmelt_VarK}
\end{center}
\end{figure*}

\subsection{Dependence on $\kappa$}

In this section, we present the melting temperature as a function of $\kappa$, the range of the attractive part of
the inter-particle potential, for $N=20$ and $N=30$ particles and a fixed value of $B$. It is important to remember
that the $\kappa$-parameter is inversely proportional to the range of the attractive part of the inter-particle
potential. Although the attractive range presented in Fig. \ref{fig:Tmelt_VarK}(a) is too small to agglomerate
particles, it is significant to change the melting temperature for small $\kappa$-values. In other words, when the
value of $\kappa$ is increased from $\kappa=4.5$ to higher values, the attraction between the particles decreases
and consequently the Coulomb repulsion increases. Consequently, the particles become more packed and the melting
temperature increases. For large $\kappa$-values shown in Fig. \ref{fig:Tmelt_VarK}(a), the attraction range is too
small either to form small clusters or to influence the melting temperature. This behavior is due to the fact that
the coupling among particles saturates for large $\kappa$-values, that is, the particles cannot be more densely
packed and therefore the value of the melting temperature becomes almost constant.

For the situation when bubbles are present (see Fig. \ref{fig:Tmelt_VarK}(b)), we find that the melting temperature
also increases with increasing $\kappa$. This general behavior is due to a decrease of the attractive part of the
inter-particle potential when $\kappa$ increases. Moreover, we observe that for small $\kappa$-values shown in Fig.
\ref{fig:Tmelt_VarK}(b), the range of the attraction is large enough to form bubbles but does not influence the
melting temperature considerably. In other words, the particles are weakly coupled in small bubbles for small values
of $\kappa$ and the small difference of $\kappa$ does not change strongly the coupling among particles and therefore
the melting temperature. On the another hand, for large $\kappa$-values in Fig. \ref{fig:Tmelt_VarK}(b), the
different attraction range changes the melting temperature due to the increase of the coupling among the particles.
In this situation, a small difference of $\kappa$ increases the packing among particles and therefore changes the
melting temperature. A similar trend  in the melting temperature is found for other clusters with $30$ particles
[Figs. \ref{fig:Tmelt_VarK}(c) and \ref{fig:Tmelt_VarK}(d) respectively]. We can see that the behavior of the
melting temperature is in general the same for the different configurations, i.e., the higher the value of $\kappa$,
the higher the coupling among particles and consequently the higher the melting temperature. However, the melting
temperature has different regimes as a function of $\kappa$ for the same configuration.

In this section, we observed clearly the different dependence of the melting temperature with respect to the
strength and the range of the attractive inter-particle potential. For different strength of the attractive
potential, the behavior of the melting temperature is determined by the configuration of the system, i.e., pure ring
configurations or bubble configurations. On the another hand, the range of the attractive potential is important to
determine the trend of the melting temperature for a specific configuration. The bubble configurations presented in
this section did not exhibit a structural transition from bubble to ring configuration when the temperature is
increased. This is a consequence of the fact that the melting temperature of the ring configurations are smaller
than that of the bubble configurations, i.e., the bubble configurations are more stable than the ring configurations
for the strength and the range of the attractive potential presented in Figs. \ref{fig:Tmelt_VarB}(b),
\ref{fig:Tmelt_VarB}(d), \ref{fig:Tmelt_VarK}(b) and \ref{fig:Tmelt_VarK}(d).

%
%

\section{Structural behavior}

\begin{figure}
\begin{center}
\includegraphics[scale=1]{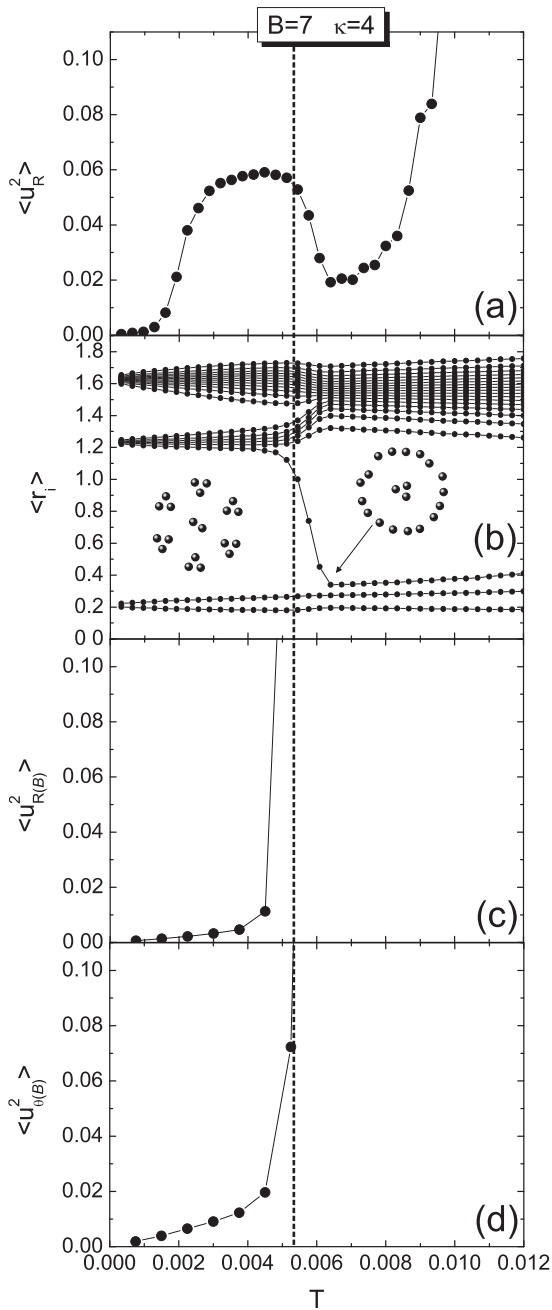}
\caption{(a) The mean radial displacement ($<u_{R}^2>$), (b) the mean distance of each closest particle from the
center of the confinement potential, (c) the mean radial displacement and (d) the angular intrashell displacement
with respect to the center of mass of the small bubble for a system with $N=20$, $B=7$ and
$\kappa=4$.}\label{fig:uRR2_B7_K4}
\end{center}
\end{figure}

Recently, colloidal systems exhibited several new and interesting features, such as re-entrant behavior
\cite{Kwinten_Re,Bubeck99,Bechinger00,Peeters00} and a hysteresis effect \cite{Reich_Hysteresis}. In this section,
we show for a specific short-range interaction that our system can present a different re-entrant effect and a
hysteresis behavior as a function of temperature.

\subsection{Re-entrant behavior}

The temperature dependence of the mean squared radial displacement $<u_{R}^2>$ is shown in Fig.
\ref{fig:uRR2_B7_K4}(a) for a system with $B=7$ and $\kappa=4$. As we can observe, when the temperature increases
from $T=0$ to $T=0.003$, the value of $<u_{R}^2>$ increases considerably until it reaches a plateau and remains
almost constant until $T=0.0053$ (dashed line). For $T>0.0053$, we observe that the value of $<u_{R}^2>$ decreases
rapidly before it increases sharply, indicating that the system melts. This re-entrant behavior was observed both in
experimental \cite{Bubeck99,Bechinger00} and theoretical studies \cite{Peeters00,Kwinten_Re} which, in the present
case, is due to an increase of the stability of the whole system caused by an increase of the symmetry of the system
configuration. In other words, the value of the mean radial displacement decreases after $T=0.0053$ as a result of
the change in the configuration of the system.

In order to confirm that the re-entrant behavior is caused by the change in the configuration, we plot in Fig.
\ref{fig:uRR2_B7_K4}(b) the position of the particles with respect to the center of the confinement potential as a
function of temperature. Specifically, in each MD step we organize the distance of each particle in such a way that
$r_1,r_2,r_3,...,r_N$ correspond to the first, second, third,..., $N$th closest particles from the center of the
cluster, respectively. After that, we calculate the average of each closest distance $<r_i>$ for each temperature
and we present it as a function of temperature in Fig. $\ref{fig:uRR2_B7_K4}$(b). As can be seen, the system remains
in the bubble configuration from $T=0$ to $T=0.0053$. We will label this configuration as ($2;6(3)^B$) which means
that there is one ring of $2$ particles and $6$ bubbles of $3$ particles [see the left inset of Fig.
\ref{fig:uRR2_B7_K4}(b)]. For $T>0.0053$, a particle leaves from the edge and goes to the center of the system. The
configuration changes from the bubble ($2;6(3)^B$) to the ring configuration ($3;17$), which means that there is a
ring of $3$ particles and another with $17$ particles. We observe from Figs. \ref{fig:uRR2_B7_K4}(a) and
\ref{fig:uRR2_B7_K4}(b) that the value of the mean radial displacement decreases when the configuration changes from
($2;6(3)^B$) to ($3;17$) and increases rapidly when the temperature approaches the melting temperature of the ring
configuration. Therefore, this change of configuration is a structural transition which stabilizes the system. This
transition occurs before the system is completely melted and is a thermally induced structural phase transition.
This interesting phenomenon was found in diverse previous studies \cite{tomecka,coupier05,wand05} but, in this case,
it is the result of the increase of disorder in small bubbles when the temperature approaches $T=0.0053$.

In order to better understand the local disorder of the particles, we present in Figs. \ref{fig:uRR2_B7_K4}(c) and
\ref{fig:uRR2_B7_K4}(d) the mean radial displacement and angular disorder of the particles of the small bubbles as a
function of temperature. In particular, we calculate the radial and angular disorder of each particle with respect
to the center of mass of the small bubble to which it belongs. Due to the rotation of the small bubble with respect
to the confinement center, the center of mass of each small bubble is calculated in each MD step. The mean locally
radial displacement is defined as
%
%
\begin{equation}\label{radialdeviation}
\displaystyle{
  \langle u_{R(B)}^{2} \rangle \!\!\equiv
  \!\!\frac{1}{N_{c}}\frac{1}{N_{sc}}\sum_{j=1}^{N_{c}}\sum_{i=1}^{N_{sc}}[\langle (r_{j}^{cm}\!\!\!-r_{i})^2 \rangle\!\!-\!\langle (r_{j}^{cm}\!\!\!-r_{i})
  \rangle ^{2}] / \rho_{B}^{2},
}
\end{equation}
where $N_c$ and $N_{sc}$ are the number of small bubbles and the number of particles in each small bubble,
respectively. $r_{j}^{cm}$ is the distance of the center of mass of the bubble from the center of the confinement
potential and $\rho_{B}$ is the average distance between the particles of the same small bubble at zero temperature.

\begin{figure}[t]
\begin{center}
\includegraphics[scale=0.85]{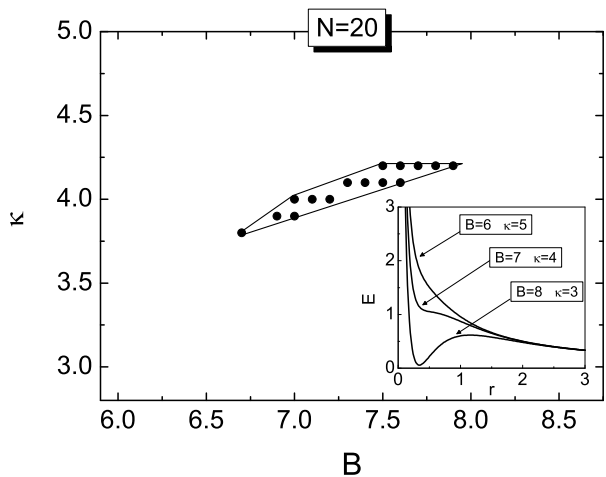}
\caption{ Phase diagram in ($B$,~$\kappa$) parameter space showing the re-entrant behavior for $N=20$ particles
(solid symbols are the calculated values). The profile of the inter-particle potential for some relevant ($B$,
$\kappa$) values is shown in the inset. }\label{fig:Phase_Diagram}
\end{center}
\end{figure}

The angular disorder in the small clusters are studied using the angular intrashell displacement calculated locally.
Previously, this property was used to calculate the angular disorder in the whole system \cite{Bedanov94,Felipe_M},
but in this case, we calculated the angular disorder in each small bubble, i.e., with respect to the center of mass
of the small agglomerate of particles which is defined as
\begin{equation}\label{angdeviation1}
\displaystyle{
  \langle u_{\theta(B)}^{2} \rangle \!\!\equiv
  \!\!\!\frac{1}{N_{c}}\frac{1}{N_{sc}}\sum_{j=1}^{N_c}\sum_{i=1}^{N_{sc}}[\langle (\varphi_{i}\!-\varphi_{i1})^{2} \rangle\!\!-\!\langle
  (\varphi_{i}\!-\varphi_{i1}) \rangle ^{2}] / (\varphi_0^{s})^2,
}
\end{equation}
where $i_1$ indicates the nearest neighbor in the same bubble and $\varphi_0^{s}=2\pi/N_{sc}$ is the average density
of particles in each small bubbles at zero temperature.

As we can see in Fig. \ref{fig:uRR2_B7_K4}(c) and \ref{fig:uRR2_B7_K4}(d), the mean radial displacement and the
angular disorder of particles increase simultaneously very dramatically in each bubble when the temperature
approaches $T=0.0053$ (dotted vertical line). This disorder permits a particle to overcome the potential barrier,
between the center particles and the edge particles, allowing to go to the center and changing the system
configuration. Notice that the local disorder properties increase rapidly before the system changes its
configuration from ($2;6(3)^B$) to ($3;17$), that is, these properties change fast because of the increase of radial
and angular disorder in the small bubble leaving initially the configuration unchanged.

\begin{figure}
\begin{center}
\includegraphics[scale=0.8]{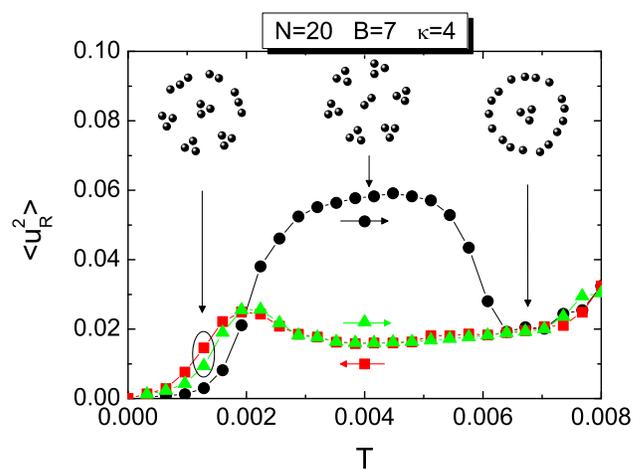}
\caption{The mean radial displacement for a cluster with $N=20$, $B=7$ and $\kappa=4$ for increasing (black ball and
green triangle symbols) and decreasing temperature (red squared symbols).}\label{fig:Loop}
\end{center}
\end{figure}

Re-entrant behavior was found in a small region of ($\kappa$ ,$B$)-space which is shown in Fig.
\ref{fig:Phase_Diagram}. We found that this interesting feature is the result of the inter-particle potential
profile which is illustrated in the inset of Fig. \ref{fig:Phase_Diagram} for some values of $B$ and $\kappa$. We
observe that the inter-particle potential for the system with $B=7$ and $\kappa=4$ is not so repulsive as that for
$B=6$ and $\kappa=5$ and it is not so attractive as for $B=8$ and $\kappa=3$. Therefore, systems which have values
of $B$ and $\kappa$ close to $7$ and $4$ respectively, are ideal to exhibit a re-entrant behavior due to the
characteristic of agglomerating particles that are sufficiently weakly bound to allow particles to overcome the
potential barrier when temperature increases, i.e., a thermally induced structural phase transition.

The re-entrant phenomenon presented for the cluster with $N=20$ particles was also observed for other values of $N$,
e.g. for $N=30$ particles. The general behavior of $<u_{R}^2>$ shown in Fig. \ref{fig:uRR2_B7_K4}(a) is also
observed for $N=30$ particles for a slightly different region in ($\kappa$, $B$)-space as well as the thermally
induced structural phase transition.

\subsection{Hysteresis behavior}


Fig. \ref{fig:Loop} shows $<u_{R}^2>$ when we decrease temperature after the system has changed its configuration to
$(3;17)$ for the same previous case, i.e., $N=20$, $B=7$ and $\kappa=4$. Specifically, the mean radial displacement
is calculated where the temperature is increased from $T=0$ to $T=0.008$ (solid circles). In $T=0.008$, the system
does not melt but reaches a different configuration from the $T=0$ configuration. After that, we decreased
temperature until $T=0$ (red square symbols) and increased it again up to $T=0.008$ (green triangle symbols). Notice
that at $T=0.008$ the system reaches a different configuration from the ground state one and that the $<u_{R}^2>$
behavior is very different when temperature is decreased indicating that the system gets stuck in a meta-stable
state. This phenomenon is very interesting because the system exhibits a hysteresis effect in the mean radial
displacement as a function of temperature when we decrease the temperature after a thermally induced phase
transition, i.e., after a change of configuration. This interesting behavior is a consequence of the fact that the
$(3;17)$ configuration is stable, although it has a larger potential energy than that of the ($2;6(3)^B$)
configuration, and is separated from the ($2;6(3)^B$) configuration by a high energy saddle point.

\begin{figure}
\centering
\includegraphics[scale=0.8]{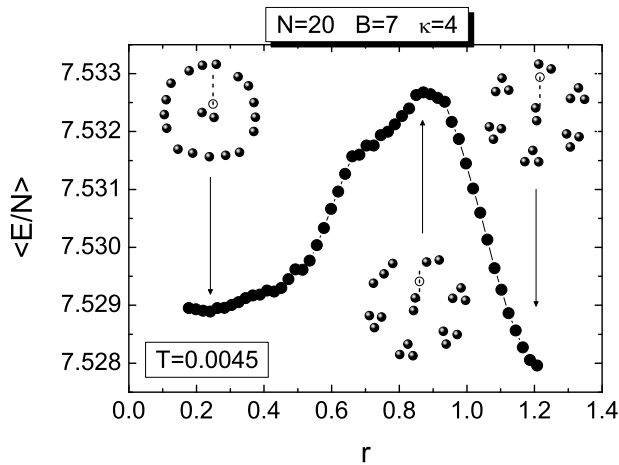}
\caption{The mean energy per particle as a function of the distance of the marked particle (particle in the inset
indicated by a open circle) from the center of the confinement potential for the cluster with $B=7$, $\kappa=4$,
$N=20$ particles and fixed temperature $T=0.0045$. The configurations at the three local energy are shown in the
inset as well as the trajectory of the marked particle (dashed line).}\label{fig:Saddle_Point}
\end{figure}

In order to visualize this saddle point in the potential energy landscape between the ($3;17$) and ($2;6(3)^B$)
configuration, we fixed one particle from the edge of the ($2;6(3)^B$) configuration at $T=0.0045$ and moved it in
the direction of the center and at the same time allowing all the other particles to relax to their equilibrium
positions. We calculated the average energy per particle of the system as a function of the position of the fixed
particle [Fig. \ref{fig:Saddle_Point}]. In this simulation we use Monte-Carlo (MC) technique in order to obtain the
ground state configuration and to make an average of the particle energy of several MC steps (around $10^7$). That
average energy per particle is presented in Fig. \ref{fig:Saddle_Point} as a function of the distance of the marked
particle (open symbol in the inset and this particle is moved along the thin dashed line) with respect to the center
of confinement for a fixed temperature $T=0.0045$.

In Fig. \ref{fig:Saddle_Point}, we observe that the average energy per particle increases sharply when the marked
particle is moved from $r=1.2$ to $r=0.9$. After that, the mean energy of the system decreases continuously until
$r\sim0.3$ and remains almost constant until $r=0.17$. The ($3;17$) configuration is found when $r=0.21$. Notice
that the ($3;17$) configuration has the second lowest energy of the whole simulation and is stable. Thus, when we
increase the temperature of the ($2;6(3)^B$) configuration such that there is enough thermal energy for a particle
to overcome the potential energy saddle point, of $T=0.007$, the system changes its configuration to ($3;17$) and is
locked into this lowest energy configuration. Consequently, the system remains in this meta-stable configuration
even when temperature is decreased down to $T=0$.

\section{Conclusion}
In this paper we investigated the dependence of the melting temperature of a system composed of classical particles,
interacting through an inter-particle potential with a short-range attractive part and a long-range repulsive part,
confined by a parabolic trap. The melting temperature showed diverse behaviors as a function of the parameters that
characterize the inter-particle interaction. In general, the melting temperature changes gradually as a function of
the strength of the short-range interaction ($B$), however, different behaviors of the melting temperature were
observed for pure ring configurations and bubble configurations as a function of $B$. The same trend of the melting
temperature was observed for different range of the attractive part of the inter-particle potential ($\kappa$) as
well as for different configurations, that is, the melting temperature increases with an increase of the $\kappa$
parameter. However, different regimes of the melting temperature appear in the same configuration as a result of a
saturation of the coupling among particles.

The mean radial displacement $<u_{R}^2>$ showed a re-entrant behavior as a function of temperature. We found that
this behavior is a consequence of a thermally induced structural phase transition which stabilizes the system before
it melts. This structural transition occurs due to the rapid increase of the local disorder in the small bubbles of
the system. This re-entrant behavior is found for a restricted set of values of the $B$ and $\kappa$ parameter which
define the attractive part of the potential. These values were shown in a phase diagram and showed that the
re-entrant behavior is a characteristic of the inter-particle potential of the system.

A hysteresis effect was observed in the structural and the dynamical behavior of the system as a function of
temperature. It was shown that this behavior is a consequence of the existence of a high energy saddle point in the
potential energy landscape between the two lowest energy configurations.

\begin{acknowledgments}
F.F.M., W.P.F. and G.A.F. were supported by the Brazilian National Research Councils: CNPq and CAPES and the
Ministry of Planning (FINEP). Part of this work was supported by the Flemish Science Foundation (FWO-Vl).
\end{acknowledgments}


\begin{thebibliography}{99}

\bibitem{Chem2} D. Philp and J. F. Stoddart, Angew. Chem. Int. Ed. Engl. {\bf35}, 1154 (1996).
\bibitem{Chang00} H. C. Chang, T. L. Lin, and G. G. Chang, Biophysical Journal {\bf78}, 2070 (2000).
\bibitem{Sear98} R. P. Sear and W. M. Gelbart, J. Chem. Phys. {\bf110}, 4582 (1999).

\bibitem{Stojkovic99} B. P. Stojkovi\'{c}, Z. G. Yu, A. L. Chernyshev, A. R. Castro Neto, and N. Gr$\phi$nbech-Jensen, Phys. Rev. B {\bf62},
4353 (1999).
\bibitem{Stojkovic00} B. P. Stojkovi\'{c}, Z. G. Yu, A. L. Chernyshev, A. R. Castro Neto, and N. Gr$\phi$nbech-Jensen, Phys. Rev. Lett. {\bf82}, 4679 (2000).
\bibitem{Bubeck99} R. Bubeck, C. Bechinger, S. Neser, and P. Leiderer, Phys. Rev. Lett. {\bf82}, 3364 (1999).
\bibitem{Leiderer98} Q. H. Wei, C. Bechinger, D. Rudhardt, and P. Leiderer, Phys. Rev. Lett. {\bf81}, 2606 (1998).
\bibitem{Schmalian00} J. Schmalian and P. G. Wolynes, Phys. Rev. Lett. {\bf85}, 836 (2000).
\bibitem{Fradkin99} E. Fradkin and S. A. Kivelson, Phys. Rev. B {\bf59}, 8065 (1999).
\bibitem{Sciortino02} F. Sciortino, Nature Mater. {\bf1}, 145 (2002).
\bibitem{Mladek06} B. M. Mladek, D. Gottwald, G. Kahl, M. Neumann, and C. N. Likos, Phys. Rev. Lett. {\bf96}, 045701 (2006).
\bibitem{Campbell05} A. I. Campbell, V. J. Anderson, J. S. van Duijneveldt, and Paul Bartlett, Phys.~Rev.~Lett. {\bf94}, 208301
(2005).
\bibitem{Lu06}P. J. Lu, J. C. Conrad, H. M. Wyss, A. B. Schofield, and D. A. Weitz, Phys.~Rev.~Lett. {\bf96}, 028306 (2006).
\bibitem{Shevchenko06} E. V. Shevchenko, D. V. Talapin, N. A. Kotov, S. O'Brien, and C. B. Murray, Nature (London) {\bf439}, 55 (2006).
\bibitem{Hoffmann06} N. Hoffmann, F. Ebert, C. N. Likos, H. Lowen, and G. Maret, Phys. Rev. Lett. {\bf97}, 078301
(2006).
\bibitem{Bedanov94} V. M. Bedanov and F. M. Peeters, Phys. Rev. B {\bf49}, 2667~(1994).
\bibitem{Kong03} M. Kong, B. Partoens, and F. M. Peeters, New J. Phys. {\bf5}, 23~(2003).
\bibitem{Peeters00}I. V. Schweigert, V. A. Schweigert, and F. M. Peeters, Phys.~Rev.~Lett. {\bf84}, 4381 (2000).

\bibitem{Kwinten_Re} K. Nelissen, B. Partoens, I. Schweigert, and F. M. Peeters, Europhys.~Lett. {\bf74}, 1046
(2006).
\bibitem{Felipe_M} W. P. Ferreira, F. F. Munarin, G. A. Farias, and F. M. Peeters, J. Phys.: Condens. Matter {\bf18},
9385 (2006).

\bibitem{Reich_InfBuble} C. J. Olson Reichhardt, C. Reichhardt, I. Martin, and A. R. Bishop, Physica D {\bf193}, 303 (2004).
\bibitem{Kwinte_Bubles} K. Nelissen, B. Partoens, and F. M. Peeters, Phys.~Rev.~E {\bf 71}, 066204 (2005).

\bibitem{Metropolis} N. Metropolis, A.~W.~Rosenbluth, M.~N.~Rosenbluth, A.~M.~Teller, and E.~Teller, J.~Chem.~Phys. {\bf21}, 1087 (1953).
\bibitem{Peeters95} V.~A.~Schweigert and F.~M.~Peeters, Phys.~Rev.~B {\bf51}, 7700 (1995).
\bibitem{lozovik85} Yu.~E.~Lozovik and V.~M.~Fartzdinov, Solid~State~Commun. {\bf 54}, 725 (1985); V.~M.~Bedanov, G.~V.~Gadiyak, and Yu.~E.~Lozovik,
Phys.~Lett. A {\bf 109}, 289 (1985).

\bibitem{Bechinger00}C. Bechinger, Q. H. Wei, and P. Leiderer, J. Phys.: Condens. Matter {\bf12}, A425 (2000).

\bibitem{Reich_Hysteresis} C. J. Olson Reichhardt, C. Reichhardt, and A. R. Bishop, Europhys.~Lett. {\bf72} (3), 444 (2005).


\bibitem{wand05} W.~P.~Ferreira, B.~Partoens, F.~M.~Peeters, and G.~A.~Farias, Phys.~Rev.~E {\bf71}, 021501 (2005).
\bibitem{tomecka} D. Tomecka, B. Partoens, and F. M. Peeters, Phys.~Rev.~E {\bf 71}, 062401 (2005).
\bibitem{coupier05} G. Coupier, C. Guthmann, Y. Noat, and M.~Saint Jean, Phys.~Rev.~E {\bf 71}, 046105 (2005).


\end{thebibliography}
\end{document}